\documentclass{sigma}
\def\ll{\label}
\def\re{\ref}
\def\c{\cite}

\def\r1{(\ref{$1})}

\def\th{\theta}
\def\ba{\begin{array}{c}}

\def\ea{\end{array}}

\def\si{\sigma}

\def\de{\delta}

\def\ov{\over}
\def\ha{{1\over 2}}

\def\l{\left}
\def\l({\left(}
\def\r){\right)}
\def\r{\right}

\def\la{\lambda}
\def\al{\alpha}

 \def\be{\begin{equation}}
\def\bc{\begin{center}}
\def\ec{\end{center}}
\def\bit{\begin{itemize}}
\def\eit{\end{itemize}}
\def\ee{\end{equation}}
\def\ed{\end{document}}
\def\bea{\begin{eqnarray}}
\def\eea{\end{eqnarray}}
\def\efr{\end{flushright}}

\begin{document}

\renewcommand{\PaperNumber}{??}

\FirstPageHeading

\ShortArticleName{Integrable higher NLS type equations}

\ArticleName{Integrable Hierarchy of Higher Nonlinear  
Schr\"odinger Type Equations }

% Names of the authors for the title of the paper
\Author{Anjan Kundu}
\AuthorNameForHeading{A. Kundu}

\Address{Saha Institute of Nuclear Physics,  
 Theory Group \& 
 Centre for Applied Mathematics \& Computational Science
, 1/AF Bidhan Nagar, Calcutta 700 064, India} % Address of First Author
\EmailD{anjan.kundu@saha.ac.in} % E-mail address of First Author
\URLaddressD{http://www.saha.ac.in/theory/anjan.kundu/} %URL address of First Author
% Address of Second Author
%\Address{$^\ddag$~Address of Second Author, Country}
%\EmailD{email@address} % E-mail address of Second Author

% In the case of the same organization, please use the following standard
%\Author{First Names LASTNAME and Second COAUTHOR}
%\AuthoqNameForHeading{F.N. Lastname and S. Coauthor}
%\Address{Address of Author(s), Country}
%\Email{email1@address, email2@address}
%\URLaddress{URL1, URL2)

\ArticleDates{Received August 14, 2006, in final form October 17, 2006;
 Published online November ??, 2006\\
Original article is available at
http://www.emis.de/journals/SIGMA/206/Paper07?/}

\Abstract{
Addition of higher   nonlinear terms to the well known
integrable   nonlinear
 Schr\"odinger  (NLS)  equations, keeping the same linear dispersion (LD)
usually makes the system nonintegrable.
 We present a systematic method through a novel
  Eckhaus-Kundu hierarchy, which  can  generate  higher 
nonlinearities  in  the NLS and derivative NLS equations
  preserving their integrability. Moreover,
similar nonlinear integrable extensions can be made again in a hierarchical
way for each of the equations in the
known integrable   NLS and derivative NLS hierarchies with higher order LD,
 without changing their LD.
}

%Please type here List of Keywords for your article separated by semicolon.

\Classification{35G20; 37C85; 35G25; 37E99} % e.g. 35A30; 81Q05
% For 2000 Mathematics Subject Classification see http://www.ams.org/msc/

\section {Introduction}
  Nonlinear  
 Schr\"odinger   equation  (NLSE):
\be
iq_t+q_{xx}-2 \sigma |q|^2q=0
\ll{nls}\ee
 as well as derivative NLSE (DNLSE) 
\be
iq_t+q_{xx}-i \alpha (|q|^2q)_x=0
\ll{dnls}\ee
arising in different physical context \c{example} ,
represent  well known integrable systems. The integrability of 
such systems as well as the stability of soliton solutions inherent to such
equations are believed to be due to a fine balance between their 
linear {\it  dispersive}  and  nonlinear {\it collapsing} terms. 
Therefore, in some physical situations,  which demand  
addition of higher nonlinear terms to (\re{nls}) or (\re{dnls})
\c{Johnson77,Benney77,example2},  
this balance  apparently is lost and  the  system turns into a
nonintegrable one, not allowing analytic solutions \c{nint}.
On the other hand, together with the well known NLSE and 
DNLSE (\re{nls}) - (\re{dnls}) there exists a tower of equations 
in their  integrable hierarchies
 corresponding to higher conserved charges, where
increasingly  higher nonlinear terms do arise. However at the same time higher
order linear dispersive terms  also  appear in these integrable 
equations, apparently to
compensate for the higher nonlinearities and for restoring the balance.
 For example, the next equation in
the  NLSE  hierarchy  (evolving with time $t_3$) has the form
\be
q_{t_3}+q_{xxx}+6\sigma^2 |q|^2q_x=0
\ll{hnls}\ee
with higher nonlinearity $|q|^2 q_x $, which however  
has   a compensating higher order linear dispersive
term: $q_{xxx}$.

Nevertheless, in physical systems quite often one needs to add higher
nonlinear terms (including nonlinear derivative terms)
 without increasing the linear  dispersion (LD) of the system.
An excellent example of such a system is the Johnson's equation
\c{Johnson77}, which was derived to solve an important hydrodynamic
problem for analyzing Stokes instability in fluid flow near the critical
value of $kh $. This equation which is uniformly valid for any $kh $ is
given in the form:  
\be
A_{t}-a_1 A_{xx}-a_2 |A|^2A+a_3 |A|^4A + ia_4 |A|^2A_x-ia_5 (|A|^2)_xA
-a_6 \theta_t A  =0,
\ll{johnson}\ee
where  $a_i $ are real numbers and 
 $ \theta_x=\delta | A|^2 $ .
%$\theta_t $ is to be determined in a consistent way.
 Evidently this 
physical system contains the same linear dispersive (LD) term $a_1 A_{xx} $ as in
the NLSE and DNLSE together with their associated  nonlinear terms, whereas 
it also has  additional 5-th order nonlinearity
along with  nonlinear terms containing derivatives (described by the terms with
coefficients $a_3, a_4 $ as well as $a_6 $). As shown in  \c{Johnson77},
 at criticality:
$kh=1.363 $ and  all $a_i	 $ have known values with $a_2=0 $.

Another example of such a physical system is the Benney's hydrodynamic
equation \c{Benney77} describing long and short wave interaction.    
Therefore it is  natural to ask whether  one can  add higher
 nonlinearity
to the original   integrable NLS and DNLS equations 
 (\re{nls})-(\re{dnls}) 
 without changing 
their LD, and at the same time   preserving their
integrability.
 Moreover, we may  enquire whether  it is possible to do this in 
 a systematic and hierarchical way.

 We
  focus on these intriguing
questions and find that there exists  
   an integrable  hierarchy 
 of Eckhaus-Kundu equations  \c{kundu84,kee},  extending  the NLS and DNLS 
equations to higher nonlinearities and thus it provides a conclusive 
  answer to the above posed
questions.  Moreover,  such a hierarchy with the addition of 
 nonlinear terms in a recursive way without changing their LD and the
integrability can also  be constructed  
for each of the equations with higher order LD in the
known integrable   NLSE and DNLSE hierarchies.

\section{Higher nonlinear NLSE and DNLSE}
It has been   shown  \c{kundu84} that,  under a
nonlinear transformation of the field $q\to Q =qe^{-i \theta }$ 
%\th dimless for both NLS,dnls
with arbitrary gauge function $\th $ the NLSE
yields  an integrable higher nonlinear equation
\be
iQ_t+Q_{xx}-2 \sigma |Q|^2Q  \ - (\th_t+\th _x ^2-i \th_{xx})
Q+2i\th _x Q_x
=0
\ll{kundu1}\ee
The DNLSE  is extended similarly under the same  field 
transformation to  
\be
iQ_t+Q_{xx}-i \al ( |Q|^2Q)_x   \ - (\th_t+\th _x ^2-i \th_{xx})
Q+\th _x(2i Q_x+\al |Q |^2Q)
=0\ll{kundu2}\ee
It is interesting to note 
that the generation of such new integrable equations 
can be   linked  to the gauge
transformation of the corresponding Lax equations : $\Phi_x=U\Phi,\
\Phi_t=V\Phi,\  $ of the NLSE and DNLSE to the new systems
$\tilde \Phi_x=\tilde U \tilde \Phi,\
\tilde \Phi_t=\tilde V \tilde \Phi, $ \ with the gauge transformed
Lax operators $\tilde U=hUh^{-1}+h_xh^{-1} $ and $
\tilde V=hVh^{-1}+h_th^{-1} $, where the gauge matrix $h\in U(1) $
 is given by
$h=e^{i\th \sigma_3}$, with arbitrary gauge field $\th
(x,t) $.
Note that though both the equations (\re{kundu1}),(\re{kundu2})  have
 been studied quite extensively  \c{clark87,sasasat95,feng01}, the
investigations  were
confined mostly to the particular functional choice for $\th $ as
$ \  \th =\delta \int^x  |Q (x')|^2 dx '$, with a real parameter $\de $.
%NLS:\de\sim 1, since Q\sim [1/L]^{1/2}%%DNLS 
%\de\sim [L], since Q\sim [1/L]
 Under this choice the NLSE (\re{nls}) is extended to  the
Eckhaus-Kundu (EK)  equation \c{kundu84,kee,shen,conte}
\be
iQ_t+Q_{xx}-2 \sigma |Q|^2Q + \delta^2  |Q|^4Q+2 i\delta 
(  |Q|^2)_xQ=0
\ll{kee}\ee
while the DNLSE (\re{dnls}) turns into a similar equation  
\be
iQ_t+Q_{xx}-i\al ( |Q|^2Q)_x  
+ {\delta \ov 2} (2\delta-\al) 
  |Q|^4Q+2 i\delta 
(  |Q|^2)_xQ=0
\ll{dkee}\ee
Note that in  both the above  equations    higher  
nonlinear terms including  derivative 
 and  5-th power  nonlinearity   are added to
the NLSE and DNLSE,  
  without changing  the  linear
 dispersive term and without spoiling  the integrability  of the system. 
It was detected  \c{kundu84} that, for the parameter choice $\al =2 \delta$
 and $\al =\delta$ , (\re{dkee})
turns into two other well known equations, e.g. Chen-Lie-Liu \c{cll} and
Gerdjikov-Ivanov equations \c{GIE}. 
 
We intend  to generalize now  this concept
  and  find  the general form  for 
additional nonlinear terms,
 for both NLSE and DNLSE, such that they can be
included without altering  the  
  linear dispersion and    the  integrability. We thus find a novel
hierarchy of EK equations which  provides
 a systematic way for  
 such  construction. We observe first that in
  equation (\ref{kundu1}) 
 the gauge function  $\th $ should be  chosen through its $x$ and $t$
derivatives as
  $\th_x=\delta |q |^2 , \ \th_t=i \delta (q^*q_x-qq^*_x) $,  for obtaining 
    higher nonlinear  NLSE  (\ref{kee}), 
while one should choose in (\ref{kundu2}),  $\th_x=\delta |q |^2 , \ \ 
\th_t=i\delta (q^*q_x-qq^*_x) +{3 \ov 2} \al \de |q |^4 $, to get 
  the  extended DNLSE (\ref{dkee}). It is important to note that,    
in both these cases  the  necessary condition 
$(\th_x)_t=(\th_t)_x $ must hold, which   
  follows here  from the  validity of the
NLSE and DNLSE themselves for the field function $q(t,x)$.

 Our idea, therefore, is to 
 widen this  choice of $\th$  for
  finding  a hierarchical rule,
 by   using the conservation relation
 $i\rho_t+I_x=0 $.  Such  integrable
systems, as is well known, possess 
 infinite number of conserved quantities in
involution, and  we can make a   systematic and  recursive choice
for
 $\th^{(n)}_x=\rho^{(n)} \ $ and $ \ \th^{(n)}_t= i I^{(n)}, \ n=0,1,2,\ldots
\ $ for constructing our hierarchy.

Interestingly, we can apply  the idea of this construction
 to each of the
higher equations in the NLSE and DNLSE hierarchies 
with higher order LD to generate a new
 integrable EK hierarchy with 
 higher
nonlinear terms, but with the same   LD. 

To remove  any possible  confusion, we should remark here 
   that the integrable EK hierarchies
 we discover here for the NLSE and DNLSE, together with  the
known  equations (\ref{kee}, \ref{dkee}) 
 as well as    the well known 
 Chen-Lie-Liu  and
Gerdjikov-Ivanov equations etc.  do not appear in the celebrated 
 classification list 
of Mikhailov et al \c{MSY}, since it excludes equations
which are related to the enlisted equations through gauge or invertible
nonlinear
transformations.  
\section {Recursive formulas and general framework for
  conservation law} 
%Hierarchy of higher nonlinear equations  for NLSE and DNLSE }
We first present  a general framework for deriving the
 required conservation rule
for both NLSE and DNLSE. We start with  the linear set of  Lax equations:
 \  $ \Phi_x=U\Phi, \ 
\Phi_t=V\Phi $ with $\Phi =(\Phi_1, \Phi_2 )
$ and  eliminate $\Phi_2 $ from both these
 equations.
 That  would yield from the first Lax equation
  a second order linear equation for $\Phi_1$:
\be 
 \Phi_{1xx}=({U_{12x} \ov U_{12}})\Phi_{1x}+(U_{11x}+U_{11}^2+U_{12}U_{21}
-\frac {U_{12x}}  {U_{12}} U_{11})\Phi_1,
\ll{Phi1} \ee
where $U_{ij}$ are matrix elements of the space-Lax operator $U$. 
Representing now  $\Phi_1=e^{U_{11}x+ \phi }$ the linear equation (\re{Phi1}) 
%$\Phi_1=e^{i\lambda \sigma^3+ \phi }$
  turns into
a Riccati equation for $ \phi_x=\nu (\la)$:
\be 
U_{12}({\nu \ov U_{12}})_x +2U_{11} \nu +\nu ^2=U_{12}U_{21},
\ll{riccatig} \ee
 which by expanding $\nu(\la)$
 in powers of $\la ^{-1}$  generates
 an infinite
set of conserved densities $c_n, n=0,1,2,\ldots$.
 On the other hand from the 
time-Lax equation we derive  
$\phi_t=V_{11}+V_{12}{ \Phi_2 \ov\Phi_1}$, 
while the space-Lax equation gives ${ \Phi_2 \ov \Phi_1}=
{ \phi_x \ov U_{12}} $, yielding 
$\phi_t=V_{11}+{ V_{12} \ov U_{12}}\phi_x$. After taking   x-derivative 
this gives the crucial relation 
\be 
\nu_t=(V_{11}+{ V_{12} \ov U_{12}}\ \nu)_x
\ll{cr}\ee
Note that we  can  derive now the conservation
 law and hence the hierarchical expressions
for the gauge field $\th$ from (\re{cr}) for  concrete systems
of NLSE and DNLSE by
inserting the corresponding expressions for the Lax matrix elements 
$U_{ij}, V_{ij}$.

\section {Higher nonlinear  hierarchy from  NLSE}
We concentrate first on   the NLSE system for deriving its integrable 
hierarchy of higher nonlinear equations with the same linear dispersive term.
 For  NLSE the conservation 
relations were  established explicitly in \c{AblowS}.  Using the known form
for
the Lax matrices   of the NLSE:
\bea & & U_{11}=-i\lambda, \ U_{12}=\sqrt \sigma q  \ , 
\ U_{21}=\sqrt \sigma q^*   \nonumber \\
& \mbox {  and } & V_{11}=-i(2 \lambda ^2+\si |q |^2) , \ V_{12}=
\sqrt \sigma(2 \la q + i q_x), \ll{UV}\eea 
we  derive  the  corresponding 
Riccati equation from  (\re{riccatig}) as 
\be 
2i\la \nu=\nu^2-\sigma |q |^2 +q({\nu \ov q})_x.
\ll{riccati1} \ee
Since $\nu $ vanishes at $|\la | \to \infty $, expanding it 
 in spectral  parameter 
$\nu =\sum_{n=0} 
 {c_n \ov (2i \lambda)^{n+1} }  $, we find 
 a recursion relation from (\re{riccati1}) for the densities of the
conserved quantities:
\be 
c_{n+1}=q({c_{n}\ov q})_x+\sum _{k=0 }^{n-1 }c_k c_{n-k-1}, \ \  \mbox{for } 
n \geq 1
\ll{recur1} \ee
  with $c_0=-\si |q|^2, \ c_1=-\si qq^*_x, \ c_2=-\si( qq^*_{xx}-\si |q |^4) $,
 etc. This gives a 
 methodical  way for evaluating  the infinite
set of  conserved
quantities $C_n=\int c_n dx , n=0,1,2, \ldots$.  Since these conserved
quantities  are related  to the NLSE, which is  
 an integrable system in the Liouville sense, all of them must be in  involution :
$\{C_n,C_m \}_{PB}=0 $ \c{AblowS,soliton}. This may  be checked  by using the
fundamental PB relation $\{q(x),q^* (y)\}_{PB}=\delta (x-y)  $ and
the boundary condition $\lim_{x \to \pm \infty}| q| \to 0 $.
 We focus now on   crucial relation (\re{cr}), which   using the relevant Lax
matrix elements for the NLSE derives the conservation relation
\be 
\nu_t=-i(\sigma |q |^2 +(2i\lambda-{q_x \ov q})\nu)_x
\ll{cl1}\ee
Expanding further $\nu(\la) $ in   parameter $\la$,  we  get the
hierarchy of  relations for the densities of conserved quantities
\be 
c_{n t}=i\left(-c_{n+1 }+ {q_x \ov q }c_n \right )_{x }, \ \ n=0, 1,2, \ldots  
\ll{cc1}\ee
where  all 
  densities $c_n$ can be  evaluated  
 from  the recurrence relation (\re{recur1}).
Choosing therefore $\rho^{(n)}=-{1 \ov \si}c_n $ and $  I^{(n)}={1 \ov \si} 
(c_{n+1 }- {q_x \ov q }c_n) $ we 
can derive finally the required  conservation law
\be 
i\rho^{(n)}_t+I^{(n)}_x=0 , n=0, 1,2, \ldots
\ll{crel}\ee
 Considering few starting values  $n=0,1, \dots$ we can easily evaluate
   their explicit forms   as
\be  \rho^{(0)}=|q|^2, \ 
I^{(0)}=-(qq^*_x- q^*q_x ), \ \rho^{(1)}=qq^*_x, \ 
I^{(1)}=(|q_x|^2-qq^*_{xx}+\sigma |q |^4 ),
\ll{crelexp}\ee
etc. Therefore we  conclude that
  we can have a  series of 
 choices for the gauge function $\th $ given by 
\bea
\th^{(n)}_x=  \de [\rho^{(n)}]&=& -{\de \ov \si}[c_n] 
, \nonumber \\  \th^{(n)}_t=\de[i I^{(n)}]&=& {\de  \ov \si} 
[i(c_{n+1 }- {q_x \ov q }c_n)]= 
 {\de  \ov \si}[i(c_{n+1 }- {Q_x \ov Q }c_n+i{\de  \ov \si}c^2_n)],
 \ll{thcn}\eea 
 for each values of  $n=0, 1,2, \ldots $, 
where $[x]=x+x^* $ (or $i(x-x^*) $) indicates nontrivial  real valued combination of $x$ together with 
convenient normalization by some constants. For example, one should evaluate
the term  appearing in the above expression as 
$[i {Q_x \ov Q }c_n]=i( {Q_x \ov Q }c_n- {Q^*_x \ov Q^* }c^*_n)$.    
Note that by inserting  this   choice  for $\th ^{(n)} \ $
in   higher order equation    
(\re{kundu1})
one can  generate  a novel hierarchy of EK equations
%\bea
%iQ_t&+&Q_{xx}-2 \sigma |Q|^2Q  \nonumber \\ &-& 
% {\de  \ov \si}[(c_{n+1 }-i {Q_x \ov Q }c_n-ic_{n x})Q+2ic_n Q_x]=0
%\ll{EKn}\eea
for $ n=0,1,2,\ldots $.
 At the bottom of this hierarchy, as we see from (\ref {crelexp}),  
 lies the simplest equation   (\ref {kee}).
However it is important to note, that since the gauge field $\theta $  in our
construction should be a real function, 
when it is chosen 
  through conserved densities $c_n$,  
which  in general can be complex valued as evident from 
 (\re{recur1}),
  {\it real} combination of its expression
must be taken with proper care, along with any 
convenient normalization.
Such a manipulation with $c_n$ 
obviously 
does not affect the conservation law (\re{cc1}), since it is a linear
equation in $c_n$.  
 For example, for $n=0$ when
 $c_0$ is real  we can take $ \th ^{(0)} _x=\de |q|^2$, however for 
 $ n=1$, when $c_1$ is complex valued, we must have $
 \th^{(1)}_x(q)=-{\de  \ov \si}[c_1] =-i  \de (q^*q_x-qq^*_x) $ to make it real.
Similarly we get for $n=2, $  $\th^{(2)}_x(q)=-
{\de  \ov \si}[c_2]=  \de (q^*q_{xx}
+qq^*_{xx}-2 \si |q|^4) $.
% Same precaution, as indicated 
%in (\re{thcn}), has also to be taken  in defining $\th ^{(n)}_t$.

The simplest and the lowest in the EK hierarchy with $n=0 
$ is given  explicitly by
(\ref{kee}).
We therefore present   the details    for  the
 next  higher nonlinear 
equation  with  $n=1$  in the EK hierarchy  
  using (\ref{crelexp}), while the other 
equations with $n \geq 2$ can be derived in a similar way.
Following the above formulation we get for $n=1$: 
\be
 \th^{(1)}_x(q)= i\de  T(q), \  
\ \th^{(1)}_t(q)= -\de(E(q)
+2\sigma
|q |^4)  
\ll{n1}\ee
where \be
 T(q)= (q^*q_x-qq^*_x), \ \ E(q)= 2 q_x^*q_x-(qq^*_{xx}+q^*q_{xx}).
 \ll{TEnls}\ee
The consistency  of (\re{n1})
 can also be checked independently by using the NLSE for the field $q$.
However, since  the equation for the  transformed 
field $q \to Q $ with $
q=Qe^{i\th (Q)}$ is of our interest,
   we have to rewrite  all the above expressions  completely 
in terms
of the  new field, by expressing the functions $\th_x(q), \th_t(q) $
 through $Q $.
 For this purpose we derive 
$q^*q_x=(Q^*e^{-i\th (Q) })\left
 ((Q_x+{i\th_x (Q)}Q)e^{i\th (Q)}\right )=(Q^*Q_x)+i  \th_x(Q) |Q |^2 , $
which  
 evaluates \bea T(q)&=&T(Q)+2i  \th ^{(1)}_x(Q) |Q |^2, \   
E(q)=E(Q)+ 4(-i \th ^{(1)}_x(Q) T(Q)+\th ^{(1)} _x(Q)^2  |Q |^2) 
\ll{TEQnls} \\
\mbox{where}& &
 T(Q) = 
(Q^*Q_x-QQ^*_x), \ E(Q)= 2 |Q_x|^2 -(Q^*Q_{xx} +QQ^*_{xx})\ll{TE1}\eea
 %\be 
resulting 
 $\th ^{(1)}_x(Q)=i \de (Q^*Q_x-QQ^*_x)-2 \de \th ^{(1)}_x(Q) |Q |^2$
%\ll{Q}\ee
which gives 
\be 
 \th ^{(1)}_x(Q)=i {\de \ov M(Q)} T(Q), \, \ \ M(Q) \equiv  1+2 \de  |Q |^2
\ll{thxQ}\ee
and similarly 
\be 
 \th ^{(1)}_t(Q)=
\de (E(Q)+2 \si |Q|^4+4(-i \th ^{(1)}_x(Q) T(Q)+\th ^{(1)} _x(Q)^2  |Q |^2))
\ll{thtQ}\ee
where $T(Q), E(Q)
$  are  as defined  in  (\re{TE1}).

As a result we  derive
 the next   higher nonlinear
equation with $ n=1$ 
in  the  EK hierarchy  from the NLSE  as
\bea
iQ_t&+&Q_{xx}-2 \sigma |Q|^2Q  \ - \de 
 ((E(Q)+2 \sigma |Q|^4+
{1 \ov M (Q)}T_x(Q))Q +2{1 \ov M (Q)}T(Q) Q_x) \nonumber \\
 &+&{\de ^2\ov M (Q)} T(Q) ({1 \ov M (Q)}(4 \de T(Q)+(|Q |^2 )_x)-3 T(Q))Q
=0
\ll{KEnks2}\eea
with $E(Q),M(Q), T(Q) $ as defined above.
Note that this is an integrable equation 
with   higher nonlinearities,
  but 
 with the same second order LD as in the
NLSE. Notice that (\ref{KEnks2}) contains an additional coupling constant $
\delta $ entering  in different powers. Therefore we may simplify this equation 
by considering  $\de$
to be small and  ignoring all terms with higher powers in $\de $, which
yields
\be 
iQ_t+Q_{xx}-2 \sigma |Q|^2Q  +2 \de 
 (- \si |Q|^4Q- Q^*(Q_x)^2 +Q_{xx}^*Q^2 )=0. \ll{simple}\ee
However it should be noted that though the simpler  equation (\ref{simple})
 as such is  a nonintegrable system, it 
 might be meaningful for physical applications and  useful
information can be extracted for it 
from its integrable variant (\ref{KEnks2}) 
 through limiting procedure.

The general form of the EK hierarchy   can  be
generated similarly in the same form  (\re{kundu1})
 where  the series of 
 gauge functions $\th^{(n)} _t, \th^{(n)} _x, n=0,1,2,\ldots $ are
chosen as (\re{thcn}).
This would construct finally  a  new 
hierarchy  of integrable  EK equations emerging  from the  NLSE   in the form
\bea
iQ_t&+&Q_{xx}-2 \sigma |Q|^2Q  \ + {\de \ov \si} (
([i(-c_{n+1 }+ {Q_x \ov Q }c_n)] -i[c_{n x }])
Q-2i[c_n] Q_x)=0,
\ll{kundu1n}\eea
  where the real-value  sign
 $[\cdot ]$ should be  properly evaluated  
 in explicit
calculations. 
 The  conserved densities   $c_n=c(Q)_n , n=0, 1,2,$ appearing in equations
(\ref{kundu1n}) are to be 
  obtained
systematically from (\re{recur1}) and then expressed consistently  in terms
of the field $Q$, as we have demonstrated in the case $n=1 $.
We stress again that   the integrable hierarchy of EK equations
(\ref{kundu1n}),
contains higher and higher   
  nonlinearities (including nonlinear dispersive terms), but  has
   the same  
second order LD  as in the original NLSE.
Note  that the density of the conserved quantities, as evident from
(\ref{recur1}),
introduces  an additional derivative resulting
the appearance of higher nonlinear dispersive terms at each higher step.
 However this 
should not be confused with the {\it linear dispersive term}
  $Q_{xx} $, which remains the
same for  the whole EK hierarchy, which in fact  is our aim.  

\subsection{LD preserving EK hierarchy  from NLSE hierarchy }
Recall that the well known  integrable  NLSE hierarchy 
is generated by the Lax pair $(U,V^{(k)})_{ NLSE}, \\   k=1,2,3, \ldots $
by the flatness condition 
$U_{t_k}-V^{(k)}_x+[U,V^{(k)}]=0 $, with the kth
equation having the   kth order LD . $k=2$ corresponds to the NLSE
(\ref{nls}), while   $k=3 $ yields the integrable equation 
(\ref{hnls}). It should be noted also that while the space-Lax operator $U $
is the same as that of the  NLSE, the time-Lax operator 
$ V^{(k)}, k=1,2,3, \ldots $ is different for different  equations in this 
hierarchy with {\it higher time} $t_k$. Nevertheless,  
 it is interesting  that, the  idea we have developed above 
for generating nonlinear
EK hierarchy from the NLSE preserving its  LD, can be successfully 
implemented 
to each of the $k$ equations in the known NLSE hierarchy. Moreover,
since the Riccati equation (\ref{riccati1})
 and consequently the recurrence relation 
(\ref{recur1}) yielding the conserved quantities depend only on the 
 Lax operator $U$, they should be the same for all equations in the
hierarchy with any
$k$. Therefore one can make the 
 same choice (\ref{thcn})
 for the gauge function $ \theta^{(n)}_x, n=0,1,\ldots $. However since the time 
evolution 
(\ref{cl1}) with respect to higher time $t_k $ would 
depend explicitly on the matrix elements  
$ V^{(k)}_{11},V^{(k)}_{12}, k=1,2,3, \ldots $
as evident from (\ref{cr}), this would be different for different $k$, leading to 
new   choices for the 
$ \theta^{(n)}_{t_k}$, obeying the consistency $(\theta^{(n)}_{x})_{t_k}=
(\theta^{(n)}_{t_k})_x. $
  Thus the kth equation in the NLSE hierarchy (with kth order LD) would
generate
EK type nonlinear hierarchy with $n=0,1,\ldots $ , preserving its linear 
(kth order)
dispersive term and the integrability.

Since we have presented above in detail the $k=2$ case, i.e the NLSE, we
give here in brief the construction  of the EK hierarchy 
 for the next equation (\ref{hnls}) with
3rd-order LD.  It is easy to check that (\ref{hnls}) can be extended
through gauge transformation to the hierarchical form:
\bea
Q_{t_3}+Q_{xxx}& +&6\sigma^2 |Q|^2Q_x+3i \theta^{(n)}_xQ_{xx}+3(i
\theta^{(n)}_{xx}-
(\theta^{(n)}_x)^2)Q_x \nonumber \\ & -& (-i\theta^{(n)}_{t_3} +3
\theta^{(n)}_{x}\theta^{(n)}_{xx}-i
\theta^{(n)}_{xxx}+i
(\theta^{(n)}_{x})^3+6 i \sigma^2 \theta^{(n)}_{x} |Q|^2   )Q  =0.
 \ll{hEKn}\eea
 For deriving now $\nu(\lambda)_{t_3} $ and consequently 
$\theta^{(n)}_{t_3} $
 we have to use , as mentioned above, the relevant elements of the corresponding
time-Lax operator $V^{(3)} $ given by \c{soliton}:
\bea
V^{(3)}_{11}&=& 4i \la^3 +(2i \la | q|^2- (qq^*_x-q^*q_x)) \sigma, \nonumber \\
V^{(3)}_{12}&=& (-4 \la^2q +2i \la  q_x-2\sigma  | q|^2q) \sqrt \sigma
. \ll{V3}\eea
Inserting consistent $\theta^{(n)}_{t_3}, \theta^{(n)}_x,   $ in 
(\ref {hEKn}) 
   we can   generate another new EK type integrable higher nonlinear hierarchy
for $n=0,1,\dots  $ with the same linear dispersive term  $Q_{xxx} $. 

We however will not derive here the general form of this hierarchy; instead     
 present  the simplest and the lowest equation  with $n=0 $.
 We find that the  corresponding  
gauge function $ \theta^{(0)}$ in this case can be   chosen as
\be 
\theta^{(0)}_{x}=\de |Q|^2, \ \ 
\theta^{(0)}_{t_3}=\de \left ( Q^*  Q_{xx} + Q Q^*_{xx}-Q^*_x  Q_{x}-3
\sigma^2 |Q|^4-3\de ^2 | Q|^6 +3i \de |Q|^2(Q^*  Q_{x} - Q Q^*_{x})\right)  
\ll{th0} \ee
Inserting (\ref{th0}) in   hierarchy (\ref{hEKn})  we reduce
it  to its lowest order equation
\bea
Q_{t_3}&+&Q_{xxx} +6\sigma^2 |Q|^2Q_x+ \de \{ 3i  |Q|^2Q_{xx}+3(i
 (|Q|^2)_{x}- \de 
|Q|^4)Q_x +(i   (|Q|^2)_{xx}-{3 \over 2}
 \de  (|Q|^4)_x)Q \nonumber \\ & +& i 
 (Q^*  Q_{xx} + Q Q^*_{xx}-Q^*_x  Q_{x}+3
\sigma^2 |Q|^4-4\de ^2 | Q|^6 +3i \de |Q|^2(Q^*  Q_{x} - Q Q^*_{x}))Q\}=0
. \ll{hEK0}\eea
Note that  this  is an integrable equation with 
higher nonlinearities up to 7-th power
in the field as well as with nonlinear dispersive terms, but having 
the same LD term $Q_{xxx}$ as in (\ref{hnls}).  
 
\section {Higher  nonlinear hierarchy from DNLSE}
The steps formulated above
 may be applied now  for finding  the nonlinear 
integrable hierarchy  from the  DNLSE
 with the same 2nd-order LD as in the original
equation.
 For this  purpose we shall use the 
  general form of the   conservation
relation presented in (\re{riccatig})-(\re{cr}), 
 but customize them for this particular case by considering 
 Lax operators \c{KN} associated with the DNLSE:     
\bea U&=& -i {\lambda ^2 \ov 4} \sigma^3+{i \ov 2} {\lambda  \sqrt \al }
  \left( \begin{array}{c}
  0\quad 
  q^*   \\
    \quad  
q   \quad  0 
          \end{array}   \right), \nonumber \\
  V&=& i ({\lambda ^4 \ov 8}-{\al \ov 4} \lambda ^2 | q|^2) \sigma^3+i
  \left( \begin{array}{c}
  0\quad 
  g^*   \\
    \quad  
g   \quad  0 
          \end{array}   \right), \ g={1 \ov 4 } \lambda \sqrt \al(-\la ^2
q+2iq_x+2\al | q|^2q)
 \ll{UVdnls}\eea    
Note however that   unlike the  NLSE  the required conservation
  relations   
 for the DNLSE
are not readily available  in the literature  in  explicit form
 and   have to be derived carefully 
 by inserting relevant expressions 
 from  (\re {UVdnls}).
This gives from (\re{Phi1}) the linear equation for $\Phi_1$ as
\be 
\Phi_{1xx}=({q^*_{x} \ov q^*})\Phi_{1x}+(-{\mu^2 \ov 16}-\mu { \al \ov 4} |q |^2+
+i {\mu \ov 4} {q^*_{x} \ov q^*})\Phi_{1}, \ \ \mu\equiv \la ^2
\ll{phi1} \ee
yielding in turn the 
  Riccati
equation  for the DNLS system:
\be 
q^*({\nu \ov q^*})_x+\nu ^2-{i \ov 2}\mu \nu +\mu { \al \ov 4}|q |^2 =0
\ll{riccati2} \ee
Using now the consistent expansion in the spectral parameter:
$\nu =i\sum_{n=0}{ c_n
 \ov \mu^n }  $, we get the
  recursion relation
\be 
{i \ov 2}c_{n+1}=q^*({c_{n}\ov q^*})_x+i\sum _{k=0 }^{n}c_k c_{n-k}, \ \ 
 \mbox{for } 
n \geq 0, \ \ c_0=-\ha \al |q|^2, 
\ll{dreccur} \ee
  which systematically generates all higher conserved densities  starting from
 \[ c_1=\al( iq^*q_x+ \ha \al |q |^4), \ \ 
c_2=2\al( q^*q_{xx}- {2 }i\al |q |^2(q^*q_x+\frac{1}{4}qq_x^*)-
\frac {\al ^2}{2}|q |^6  ) , \ \ etc. \]
 The   infinite set of conserved quantities 
 $C_n=\int c_n dx , n=0,1, \ldots$ for the DNLSE, which 
is a completely  integrable system,  must
be in involution. This   can  be checked directly by using its
fundamental PB structure: $\{q(x),q^*(y)\}_{PB}=\de_x(x-y) $ and the vanishing
boundary condition for the field: $|q | \to 0 $,  at $x \to \pm \infty $. 
Using further the Lax matrix elements 
(\re{UVdnls}) we derive from (\re{cr}) the crucial conservation law 
\be 
\nu_t=\left (-i \mu { \al \ov 4} |q |^2 + (-{\mu \ov 2}- i {q^*_x \ov q^*}+ \al |q|^2)\nu\right)_x. 
\ll{cldnls}\ee
Expanding $\nu(\mu) $ through spectral parameter
 $\mu $ yields the important  relation 
\be 
c_{n t}=\left(-\ha c_{n+1 }+(-i {q^*_x \ov q^* }+\al  |q|^2)c_n \right )_{x },
\ \ n=0, 1, \ldots, \ \ c_0=- \ha \al |q|^2.  
\ll{cc2}\ee
For deriving therefore the higher nonlinear  hierarchy from the  DNLSE  
we can use the general form (\re{kundu2}) with the choice for 
 the gauge function 
\bea
\th^{(n)}_x=\de [\rho^{(n)}]&=&
{\de \ov \al} [c_{n}]
, \nonumber \\ \th^{(n)}_t=[iI^{(n)}]&=&{\de \ov \al}[
-\ha c_{n+1 }+(-i {q^*_x \ov q^* }+ \al  |q|^2)c_n]\nonumber \\&=&
{\de \ov \al}[
-\ha c_{n+1 }+(-i {Q^*_x \ov Q^* }+ \al  |Q|^2)c_n-{\de \ov \al}c^2_n  ]
, \ n=0,1, 2, \ldots ,
\ll{gaugedns}\eea
where the  real-value  sign $[\cdot ] $ has the same 
 meaning as explained above. For example, for $n=0$ with real $c_0$ we 
 have $\th^{(0)}_x=\de |q|^2$, while for $n=1$ with complex valued
 $c_1$
 we must take 
 $\th^{(1)}_x=\de (i(q^*q_x-qq^*_x) + \al |q |^4)$, etc. with
 the similar reason holding also  
for the choice of $\th ^{(n)} _t$.

This derives finally the new  hierarchy  of integrable 
 higher nonlinear  equations from the  DNLSE   as
\bea
iQ_t&+&Q_{xx}-i \al ( |Q|^2Q)_x  \nonumber \\ &-& {\de \ov \al} 
\left (([-\ha c_{n+1 }-i {Q^*_x \ov Q^* }c_n] -i 
[ c_{n x}])
Q-2i[c_n] Q_x\right)
=0, 
\ll{kundu2n}\eea
where again all real-valued  expressions 
 should be  evaluated carefully and 
    $c_n=c(Q)_n$ are to be  
 obtained  recursively  from  (\re {dreccur}),  expressed explicitly
 through 
transformed field $Q$. Thus the tower of equations (\ref{kundu2n})
with $n=0,1,2, \dots $ would represent  a novel
 EK hierarchy, at the bottom of which 
 with $n=0$ giving  \[\th _x^{(0)}=\de |q|^2, \
\th _t^{(0)}= \de (i(q^*q_x- qq^*_x )+ {3 \ov 2 }\al |q |^4), \]
 lies equation 
 (\ref{dkee}).

It is worth noting that in analogy with the known reductions \c{kundu84},
The parameter choice $\al =2 \de $ in (\re{kundu2n}) would yield a new
 Chen-Lie-Liu type  hierarchy,
 while $\al = \de $ would generate another 
Gerdjikov-Ivanov type
hierarchy.  All these equations however have the same LD given by $ Q_{xx}$

For demonstrating the novelty of hierarchy  (\ref{kundu2n}),
we take up $n=1$, giving the next new equation with  
\be  \th^{(1)}_x(q)= \de (iT(q) + \al  |q |^4),
\ \th^{(1)}_t(q)=\de 
(E(q)+3i\al T(q)|q |^2+2\al ^ 2|q |^6 ),
\ll{dcrelexp}\ee
where 
\[ T(q)= 
q^*q_x-qq^*_x 
,\ \  E(q)= 2|q_x|^2-
(q^*q_{xx}+qq^*_{xx}).
\]
  For further application, as  was performed also in the NLSE case, we have to
 express the gauge fields $\th^{(1)}_x (q), \ \th^{(1)}_t (q)  $ given by 
(\re{dcrelexp})  in terms of the transformed field $Q $ and
rewrite 
 \[T(q)= T(Q) +2i\th^{(1)}_x |
Q|^2 \ \mbox{ and}
   \ E(q)=E(Q)+ 4(-i \th ^{(1)}_x(Q) T(Q)+\th ^{(1)} _x(Q)^2  |Q |^2). \]
Note that though the expressions for $  \th^{(1)}_x(q)$ and $ \th^{(1)}_t(q)$
 differ for  the  cases of NLSE and DNLSE, the entries $T(q)$ and $ E(q)$
appearing in them have  the same form  for both these cases. 
Using the above result 
after some algebra  we arrive  at the required  expressions
\begin{gather}  \th^{(1)}_x(Q)= {\de \ov M(Q)} (iT(Q) + \al |Q |^4), \ 
\nonumber \\  
\th^{(1)}_t(Q)=\de (E(Q)+3i \al T(Q)|Q|^2-6\al \th^{(1)}_x(Q)|Q |^4+
 2 \al ^2|Q |^6+ 4(-i \th ^{(1)}_x(Q) T(Q)+\th ^{(1)} _x(Q)^2  |Q |^2)) 
\ll{dcrelQ1}\end{gather}
 where \be M(Q)=1-2 \de |Q |^2, \ \ T(Q)=
Q^*Q_x-QQ^*_x \ \ \mbox{and} \  E(Q)= 2|Q_x|^2-
(Q^*Q_{xx}+QQ^*_{xx})  \ll{TEQ}\ee
By inserting   (\re{dcrelQ1}) 
with (\re{TEQ}) in the general form (\re{kundu2}) we derive finally 
for $n=1$ the next higher nonlinear 
 equation    in  the EK hierarchy linked to the   DNLSE as
\bea
iQ_t&+&Q_{xx}-i \al ( |Q|^2Q)_x   
 - (\th^{(1)}_t+(\th ^{(1)} _x) ^2-i \th ^{(1)}_{xx})
Q+\th ^{(1)}_x(2i Q_x+\al |Q |^2Q)=0
\ll{kEd2}\eea
where expressions for $\th^{(1)}_{t}(Q)$ and $ \th^{(1)}_{x}(Q)$
  expressed
through $ T(Q), E(Q) $ as  in 
 (\re{dcrelQ1}-\re{TEQ}) together with 
$ \th^{(1)}_{xx}(Q)$ by extracting the $x$-derivative 
 are to be inserted  in equation  (\ref{kEd2}) 
to get its  explicit higher 
nonlinear  form. As we see this integrable 
equation contains additional  higher nonlinear terms 
upto $ 7th$ power nonlinearity  
together with many nonlinear dispersive terms, though with the same
second-order LD: $Q_{xx} $ as in the original DNLSE. 
. 
\subsection{LD preserving EK hierarchy  from DNLSE hierarchy }
The integrable hierarchy of the DNLSE 
is generated similar to the NLSE by the associated 
 Lax pair $(U,V^{(k)})_{ DNLSE}, k=1,2,3, \ldots $. 
The space-Lax operator $U $
corresponds to the   DNLSE (\ref{UVdnls}), while  the time-Lax operator 
$ V^{(k)}, k=1,2,3, \ldots $ is different for different 
{\it higher time} $t_k$, yielding through the flatness condition
 the  k-th
equation in this hierarchy with  k-th order  LD.
   $k=2$   corresponds to the
DNLSE
(\ref{dnls}), while   $k=3 $ yields the next integrable equation \c{KN} 
\be
q_{t_3}-q_{xxx}+3i \sigma (|q|^2q_x)_x+{3 \ov 2 }\sigma ^2(|q |^4q)_x=0
\ll{hdnls}\ee
 The above scheme  we have  implemented to the DNLSE (with $k=2 $) 
 can also be applied 
to each of the $k>2$ equations 
for generating  integrable
EK type hierarchy  preserving its  kth order LD.  Moreover,
 the Riccati equation (\ref{riccati2}),
  the determining  relation 
(\ref{dreccur}) for  the conserved quantities
and consequently  the choice 
 for the gauge function $ \theta^{(n)}_x, n=0,1,\ldots $ as given in 
(\ref{gaugedns}), would be the
  same for all equations in this
hierarchy, since they all depend  on the 
 space-Lax operator $U$ only.
 However  the relations  like 
(\ref{cldnls}, \ref{cc2}) and consequently 
the consistent expression for $ \theta^{(n)}_{t_k}$, which 
are linked  to the time  evolution should be calculated using the  
 matrix elements of the time-Lax operator   
$ V^{(k)}, k=1,2,3, \ldots $.

  Thus we can construct in principle 
higher nonlinear hierarchy with $n=0,1,\ldots $, from any of 
the kth equation in the DNLSE hierarchy, 
 preserving its 
  kth order LD 
 and the integrability. In practice however each case with higher $k$ would
be more complicated due to more complicated structure 
of its time-Lax operator $V^{(k)} $.
Since we have presented  in detail the DNLSE case given by $k=2$, we
report  here in brief the construction   
 from the next equation (\ref{hdnls}) obtained for  $k=3 $, which 
yields
\bea
Q_{t_3}&-&Q_{xxx} +3i\sigma( |Q|^2Q_x)_x+{3 \ov 2 }\sigma ^2 ((|Q |^4Q )_x
+
i \theta^{(n)}_x|Q |^4Q)  \nonumber \\ & -&
3i\theta^{(n)}_x Q_{xx}-3(i
\theta^{(n)}_{xx}-
(\theta^{(n)}_x)^2)Q_x -3\sigma(( |Q|^2 \theta^{(n)}_x Q)_x+\theta^{(n)}_x 
|Q|^2 (Q_x-i \theta^{(n)}_xQ)) 
\nonumber \\ & +& (i\theta^{(n)}_{t_3} +3
\theta^{(n)}_{x}\theta^{(n)}_{xx}-i
\theta^{(n)}_{xxx}+i
(\theta^{(n)}_{x})^3)Q  =0.
 \ll{hdEKn}\eea
By inserting $\theta^{(n)}_{x} $ from (\ref{gaugedns}) 
and deriving the consistent
$\theta^{(n)}_{t_3} $ we can generate from 
(\ref{hdEKn})
the EK type integrable hierarchy with  $n=0,1,\dots  $,
 preserving  its
3rd-order LD.

The simplest and the lowest equation of this hierarchy  with $n=0 $
 may  be given explicitly  by 
the choice of the gauge function: 
\begin{gather}
%\bea 
\theta^{(0)}_{x}=\de |Q|^2, \nonumber \\  
\theta^{(0)}_{t_3}=\de \left ( Q^*  Q_{xx} + Q Q^*_{xx}-Q^*_x  Q_{x}
+3i(\alpha+\delta ) |Q|^2(Q^*  Q_{x} - Q Q^*_{x})
-3(2\de \alpha + \de ^2-{1 \ov 2} \alpha ^2) | Q|^6 \right ),  
\ll{th0d}
\end{gather}
% \eea
consistency of which can be checked by using the originating equation  
 (\ref{hdnls}). Inserting (\ref{th0d}) in (\ref {hdEKn}) one would clearly get 
an integrable equation having 
nonlinearity up to 7-th order in the field including its derivatives, but
with the linear dispersive term  $Q_{xxx} $ as in (\ref{hdnls}).
\section {Concluding remarks}
Our construction of  new EK type hierarchies  of integrable 
equations with 
 higher and higher nonlinearities, extending 
the NLSE and
the DNLSE  
   demonstrates clearly    that suitable combinations of 
nonlinear terms  may 
 be added to the  original
integrable equations without changing their linear dispersion relation,
and at the same time preserving their  integrability. 

Moreover, one can apply the whole procedure  
to each  of the $k=3,4, \ldots$  equations in the known
 integrable   NLSE and DNLSE hierarchies  
, containing  $k$th order linear dispersion.
Each of such equations therefore   would  generate 
its   new integrable EK hierarchy with higher nonlinear terms, but with the 
same linear dispersive term.

It may be  recalled that the generalized higher NLSE appearing
  in
many physical problems \c{Johnson77,Benney77, example2}, which contains higher
 nonlinear terms together with the same second-order linear dispersive term,
 can be reduced  
for particular parameter
choices    
to   integrable  equations (\re{kee}) or (\re{dkee}).
 Therefore we may hope
that   equations with 
 hopelessly complicated 
nonlinear
terms, arising in many other physical situations,
 could also be reducible to some of the higher nonlinear 
 equations in the rich integrable EK hierarchies found here and hence could
be exactly solved, at least for certain parameter choice or for some
limiting values.  
The method presented here for constructing new
integrable 
  nonlinear 
hierarchies with unchanged linear  dispersion  is general enough to be
  applicable  to other 
 integrable PDEs with  complex fields , e.g. mixed
DNLS, complex mKdV etc. and even to vector models like Manakov model. 

%\vskip .5cm

%{\small The author expresses  thanks to CAMCS (SINP) for financial support }
\LastPageEnding

\end{document}